# Efficient single-pass third-harmonic generation at 520 nm for pumping doubly-resonant OPO

Kong ZHANG [1, 2] and Junmin WANG[1, 2, 3, *]

*[1] State Key Laboratory of Quantum Optics and Quantum Optics Devices, Shanxi University,*

*Tai Yuan 030006, Shan Xi Province, People's Republic of China*

*[2] Institute of Opto-Electronics, Shanxi University,*

*Tai Yuan 030006, Shan Xi Province, People's Republic of China*

*[3] Collaborative Innovation Center of Extreme Optics, Shanxi University,*

*Tai Yuan 030006, Shan Xi Province, People's Republic of China*

** Corresponding author, email: wwjjmm@sxu.edu.cn*

# Efficient single-pass third-harmonic generation at 520 nm for pumping doubly-resonant OPO

A ~545 mW single-frequency tunable 520 nm green laser has been demonstrated using a periodically-poled potassium titanyl phosphate (PPKTP) bulk crystal based on single-pass third-harmonic generation (THG) of a 1560 nm laser via single-pass second-harmonic generation (SHG) followed by single-pass sum-frequency generation (SFG). In single-pass SHG, two cascaded periodically-poled magnesium-oxide-doped lithium niobate (PPMgO:LN) crystals were used, and ~3.5 W 780.25 nm doubled laser output is produced, corresponding to maximum doubling efficiency of 26.8%. The system can provide a pump source (520 nm) for an optical parametric oscillator for two-color entangled continuous-variable optical field generation at 1560 and 780 nm and two-color local oscillators for homodyne detection.

Keywords: Third-harmonic generation (THG), Second-harmonic generation (SHG), Sum-frequency generation (SFG), PPMgO:LN crystal, PPKTP crystal, PPMgO:sLT crystal

## 1. Introduction

Quantum entanglement is one of the most important potential resources for use in quantum communication (QC) and quantum information processing (QIP). A number of QC and QIP protocols, including quantum teleportation, quantum dense coding, and quantum cryptography, all benefit from a well-established quantum entanglement source. In long-distance QC and distributed QIP, the 1.5 μm low-loss band of telecom optical fibers is the ideal transmission channel. In addition, the 780 nm transition of rubidium (Rb) atoms is widely used in applications such as the cooling and manipulation of Rb atoms [1, 2], atomic frequency standards [3], quantum information storage [4], and atomic gravimeters [5]. Therefore, preparation of continuous-variable two-color entangled optical fields at 780 nm and 1.5 μm will provide an important quantum source for implementation of long-distance QC and distributed QIP systems. We aim to use a 520 nm laser to pump an optical parametric oscillator (OPO) or an optical parametric amplifier (OPA) to generate continuous-variable two-color entangled optical fields at 1560 nm and 780 nm [6, 7]. Concerning the two-color entanglement many papers have already successfully been reported in similar system [8, 9].

Many methods can be used to produce a ~ 520 nm single-frequency green laser. Foremost among these, argon ion lasers can yield coherent radiation at both 488 nm and 514.5 nm. In addition, solid-state lasers have been widely used. By using laser diodes to pump different gain media, it is possible to produce green light at specific wavelengths. For example, Prof. Yongmin Li's group at the authors' institute implemented a single-frequency continuous-wave 526.5 nm source using a laser-diode pumped and intra-cavity frequency doubled 1053 nm Nd:YLF (Nd-doped yttrium lithium fluoride) laser with a green output of more than 500 mW [10]. Laser diode-pumped and frequency doubled 1064 nm Nd:$YVO_4$ laser can yield a 532 nm green laser. When compared with the argon ion laser, the laser diode-pumped system can provide a more efficient and stable output. Nd:YAP (Nd-doped yttrium aluminum perovskite) has been used to prepare a watt-level 540 nm laser [11]. Recently, based on the use of an ytterbium-doped fiber laser (1030–1080 nm) and a fiber amplifier, green lasers have been generated by efficient frequency doubling using periodically-poled nonlinear crystals. As an example, Quantel Inc has already marketed a green laser product of this type with watt-level output and wavelength in the 515~540 nm range. And Nichia Inc has realized green laser diodes which are available at 100 mW with single mode and 1W with multimode operations.

An alternative way to achieve a single-frequency green laser is third-harmonic generation (THG) of a 1.5 μm telecommunications laser. Figure 1 shows the light quanta in the nonlinear frequency conversion processes for THG, second-harmonic generation (SHG), and sum-frequency generation (SFG). Figure 1 (a) shows the quanta for the general THG process, while (b) and (c) show two THG implementation methods via SHG followed by SFG. Recently, we experimentally applied the method shown in Figure 1 (b) to yield a single-frequency 520 nm green laser using cavity-enhanced SHG followed by singly-resonant cavity-enhanced SFG [12]. The 1560 nm laser beam is split into two parts: one part serves as a fundamental beam for SFG, while the other is used for cavity-enhanced SHG to 780 nm. The 1560 nm and 780 nm beams are input into the singly-resonant SFG cavity via a dichromatic mirror to produce the THG laser. In this case, the residual 1560 nm laser power from the SHG process obviously does not take part in the SFG. Additionally, this system is quite complex, because the two resonators must be actively locked. In 2011, Vasilyev *et al.* [13] reported a single-pass SFG configuration based on a periodically-poled stoichiometric lithium tantalate (PPsLT) crystal, and achieved 522 nm lasing by single-pass SFG using a 1566 nm laser and its frequency-doubled output. And in 2016, Philippe *et al.* [14] achieved 290 mW of 514 nm from 1542 nm by using a single-pass SFG configuration.

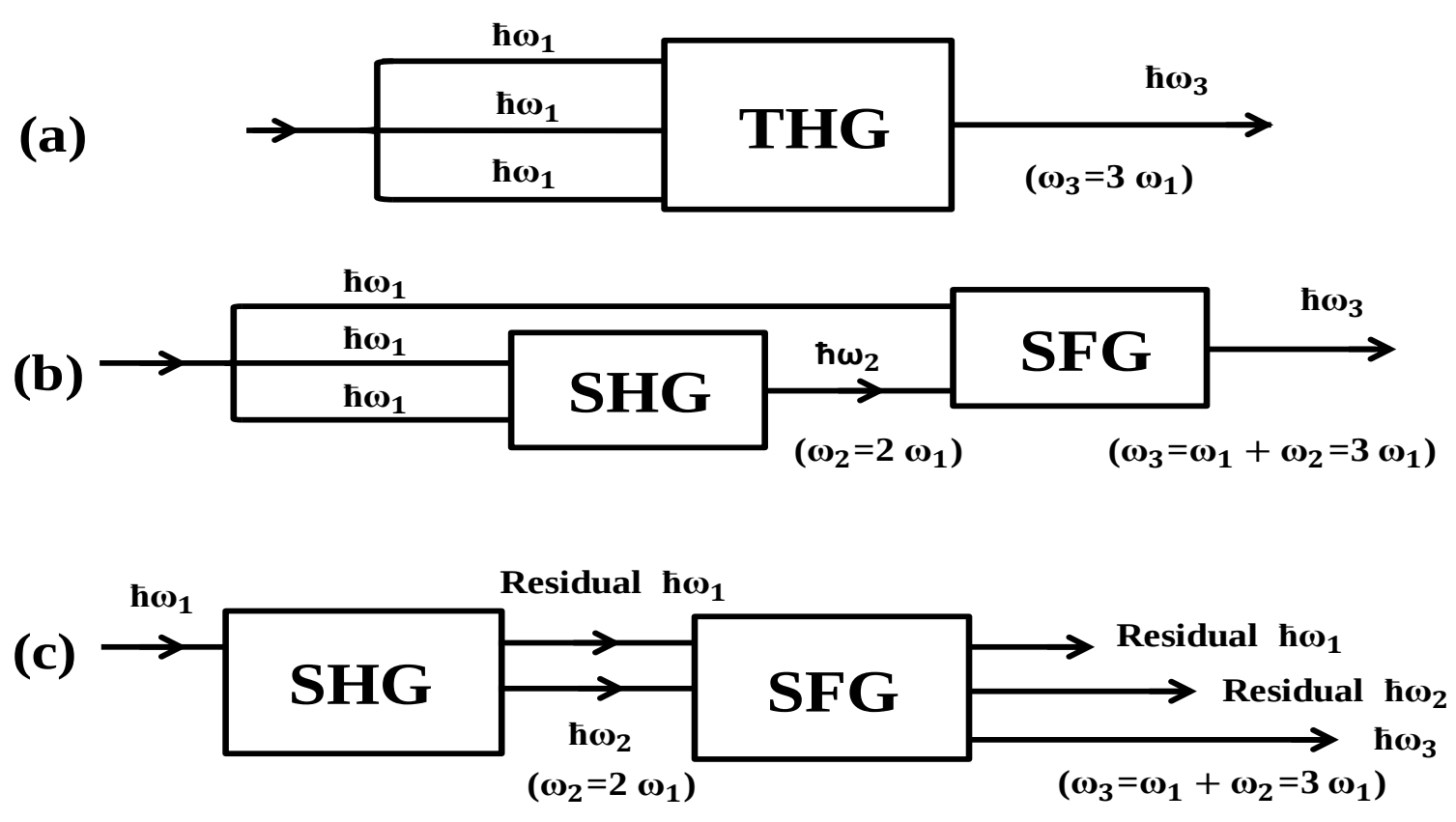

**Figure 1** (a) Light quanta in the general third-harmonic generation (THG) process. (b) and (c) Two THG implementation methods via second-harmonic generation (SHG) followed by sum-frequency generation (SFG).

Here, by following this technical line to simplify the entire laser system, we experimentally implement the method shown in Figure 1 (c). Benefit from the efficient technique of frequency doubling from 1560 nm to 780 nm [15, 16, 17] and periodically-poled crystals, we use the single-pass configuration for both the SHG and SFG processes with a 1560 nm master-oscillator power amplifier. Because of a moderate-power single-frequency THG output can still be expected when using periodically-poled nonlinear crystals and a higher 1560 nm input power, so the resonator is no longer needed. In this way, the tunable frequency range will be much larger than that in the cavity-enhanced SHG and SFG case mentioned above. Simultaneously, the 780 nm laser corresponds to the Rb D2 line, and thus an optical frequency standard can be realized by locking to the atomic absorption line to attain stable frequency characteristics for the entire laser system. Particularly, in 2011 R. Geiger *et al.* [18] detected inertial effects with airborne matter-wave interferometry in similar system. The 520 nm laser thus offers a source for generation of the continuous-variable two-color entangled optical fields at 1560 nm and 780 nm by using of OPO or OPA. In our study, we need not only a moderate ~ 520 nm single-frequency green laser to pump the OPO or OPA, but also require the corresponding 1560 nm and 780 nm laser beams to act as local oscillators for implementation of homodyne detection to verify the two-color entanglement.

This scheme has also been widely used in the generation of ultraviolet and violet lasers. Recently, Hussain *et al.* [19] produced 190 mW of 369.53 nm ultraviolet pulses by single-pass THG to $Yb^+$ fast quantum-logic. Prof. Chen's group demonstrated a 19.3 W 355 nm nanosecond laser output produced by single-pass THG using the configuration shown in Fig. 1(c) with a $K_3B_6O_{10}Br$ crystal [20].

## 2. Experimental setup for single-pass THG

The experimental scheme is shown schematically in Figure 2. A compact external-cavity diode laser (ECDL) operating at 1560.50 nm serves as the seed laser, which has a line-width of ~ 200 kHz and an output of 15 mW. The 1560.50 nm laser beam is injected into the erbium-doped fiber amplifier (EDFA). The EDFA has a narrow line-width option, which means that the fundamental light will not be obviously broadened, and we can obtain approximately 14 W of 1560.50 nm output. Single-pass cascaded 25-mm-long periodically poled magnesium-oxide-doped lithium niobate (PPMgO:LN) crystals (HC Photonics; crystal dimensions of 25.0 mm×3.4 mm×1.0 mm and 25.0 mm×3.2 mm×1.0 mm; poling period of 19.48 µm; type 0 matching; both ends of the PPMgO:LN crystals have flat surfaces with anti-reflection coatings for both the fundamental and doubled laser, and the residual reflectivity R < 0.2%) are used as the SHG crystals. The matching lens has a focal length $f$ = 50 mm with an anti-reflection coating for both the fundamental laser and doubled laser.

Subsequently, the doubled laser beam and the residual fundamental laser beam single pass through a SFG crystal to yield a 520.17 nm output. The crystal used in the SFG is a periodically-poled potassium titanyl phosphate (PPKTP) crystal or a periodically-poled magnesium-oxide-doped stoichiometric lithium tantalate (PPMgO:sLT) crystal. The matching lens has $f$ = 50 mm with an anti-reflection coating for both the fundamental and doubled laser frequencies. The optical isolator is used to restrain the laser feedback, thus ensuring the stability of the EDFA. The half-wave plate (λ/2) and the polarizing beam splitter (PBS) cube are used to control the 1560.50 nm laser power, while also transforming the polarization of the fundamental wave laser to the s polarization to meet the requirements of the frequency doubling process. During the propagation between the crystals the phase mismatch between the SH and the pump waves is accumulated, and may increase the back energy conversion from the SH signal to the pump in the second PPMgO:LN crystal. The glass plate (30mm×30mm×5mm; the insert loss of 1560nm and 780nm are 1.8% and 3.2%) is used to change the phase and reduces the back energy conversion [21], so we can yield more doubled laser finally. After SFG, a lens with anti-reflection coatings for the fundamental, doubled and tripled lasers is used to collimate the output beam. The 520 nm green light is then separated from the SFG output by using two dichroic mirrors to remove the residual 1560 nm and 780 nm components.

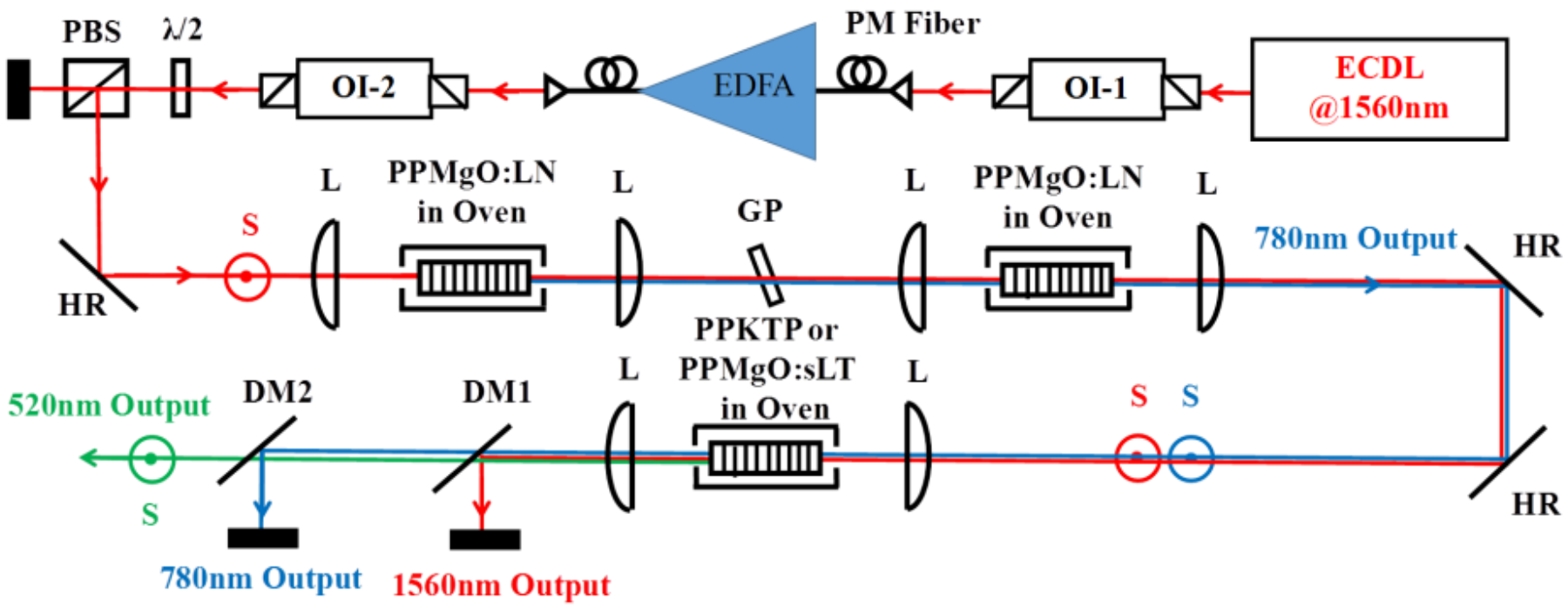

**Figure 2** Schematic diagram of setup for single-pass THG via single-pass SHG followed by SFG. ECDL: external-cavity diode laser; EDFA: erbium-doped fiber amplifier; OI: optical isolator; λ/2: half-wave plate; PBS: polarizing beam splitter cube; L: lens; GP: glass plate; DM: dichroic mirror. The 520 nm, 780 nm and 1560 nm beams are all s-polarized.

## 3. Single-pass SHG with two cascaded PPMgO:LN crystals

When one 25-mm-long PPMgO:LN bulk crystal is used for SHG, we can obtain 2 W of 780.25 nm laser output for an input of 13.15 W of 1560.50 nm fundamental laser. To improve the SHG output, we use a configuration based on two cascaded PPMgO:LN crystals, which increases the equivalent crystal length to 50 mm. The system consists of two stages. In the first stage, the 1560.50 nm laser is focused using a matching lens with $f$ = 50 mm, and the waist spot radius is ~ 35 μm, which is very close to the value determined by the optimum B-K focus factor $\xi$ = 2.84 [22]. In the second stage, both the residual fundamental beam and the frequency doubled beam are refocused at the center of the second PPMgO:LN crystal using a lens with $f$ = 50 mm, and we are also able to maintain desirable mode matching of the residual fundamental beam and the doubled beam. Subsequently, the fundamental beam and the doubled beam are collimated again. Additionally, the PPMgO:LN crystals are placed in crystal ovens, which are made from red copper and are precisely stabilized using a temperature controller (Newport Corp., Model 350B). We can then achieve the optimized phase matching by adjusting the temperatures of the crystals. When the fundamental light power is 13.15 W, the optimized matching temperatures for the cascaded crystals are 81.3 ℃ and 79.6 ℃, respectively. Figure 3 shows the experimental results (the red squares represent experimental data for one PPMgO:LN crystal, and the red circles represent data for two cascaded crystals). We can yield 3.53 W of 780.25 nm laser light, where the maximum doubling efficiency is 26.8%. We can expect that if we use longer single crystal, for example, 40 mm or 50 mm, the results should be better.

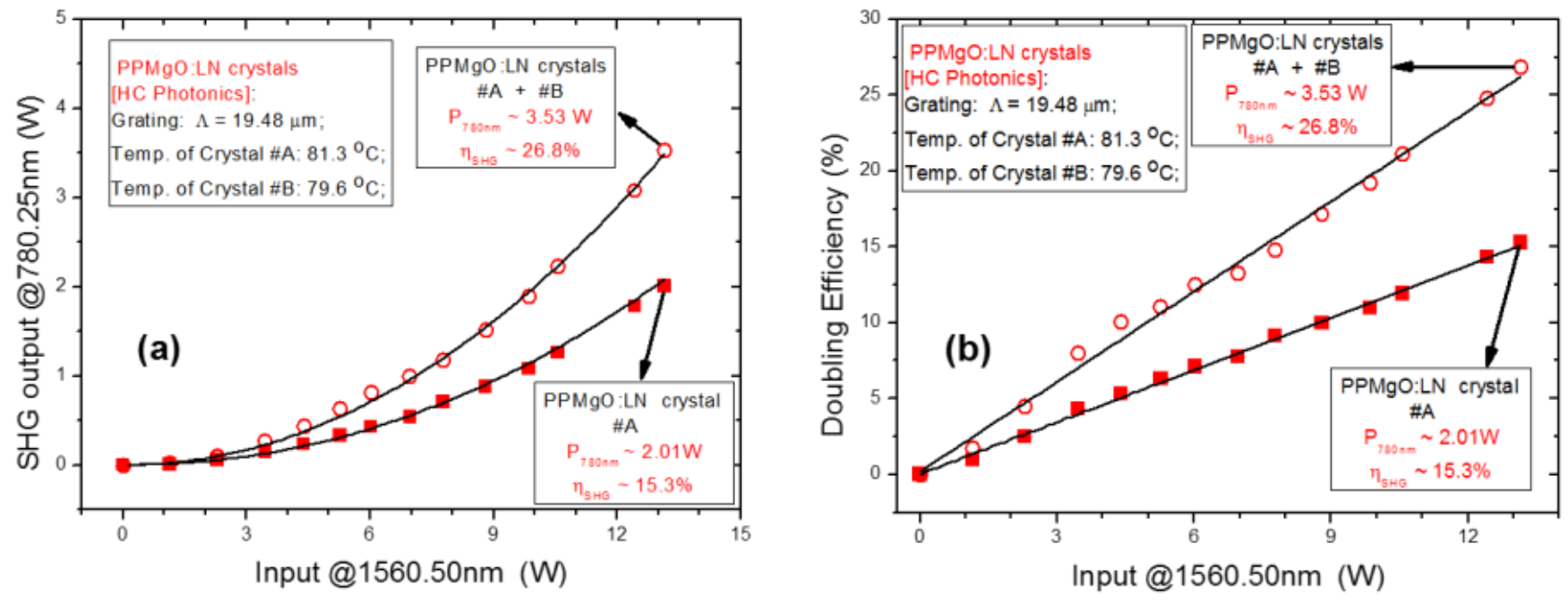

**Figure 3** (a) Experimental data for single-pass SHG using two cascaded PPMgO:LN crystals. Squares represent experimental data when using one crystal; circles represent experimental data for two cascaded crystals. (b) Nonlinear conversion efficiency of single-pass SHG using two cascaded PPMgO:LN crystals. Squares represent experimental data when using one crystal; circles represent experimental data for two cascaded crystals.

For the subsequent single-pass SFG, we need more 780.25 nm laser light than 1560.50 nm light, if we take the optical power consumption ratio between the 1560 nm and 780 nm beams in SFG process. Therefore, improvement of the doubling efficiency is an important issue. In 2011, Kumar *et al.* [23] produced 55% conversion efficiency and multi-watt output power at 532 nm using a novel scheme based on three cascaded crystals. In 2016, they subsequently demonstrated a multi-crystal scheme operating at 532 nm with four beta-phase barium borate (BBO) crystals [24]. The generated SHG light and the available fundamental radiation after each stage are collimated and refocused at the centers of the crystals. Approximately 37.3 mW of 266 nm ultraviolet light was produced using 9.2 W of 532 nm light. In addition, the interference between the SHG light beams generated in the two crystals can also improve the doubling efficiency. In 2014, Hansen *et al.* [25] demonstrated an enhancement factor of 2.3 over the single-crystal case in a cascaded two-crystal scheme for SHG.

We checked the frequency tunability of the 780.25 nm output by measuring the absorption spectra of Rb atomic vapor cells, while the 1560.50 nm fundamental laser's frequency is scanned linearly. Figure 4 shows the Doppler-broadened absorption spectra of the $5S_{1/2}$ - $5P_{3/2}$ transition (D2 line) for $^{87}$Rb and $^{85}$Rb atoms. This indicates that the continuously tunable range of the doubled laser at 780.25 nm is at least 10 GHz. It is only limited by the 1560 nm seed laser's tunability. In previous work, we demonstrated a continuous-wave 780 nm laser with watt-level output power using cavity-enhanced SHG [26]. But compared with this single-pass configuration, the continuously tunable range for cavity-enhanced SHG should be a bit small.

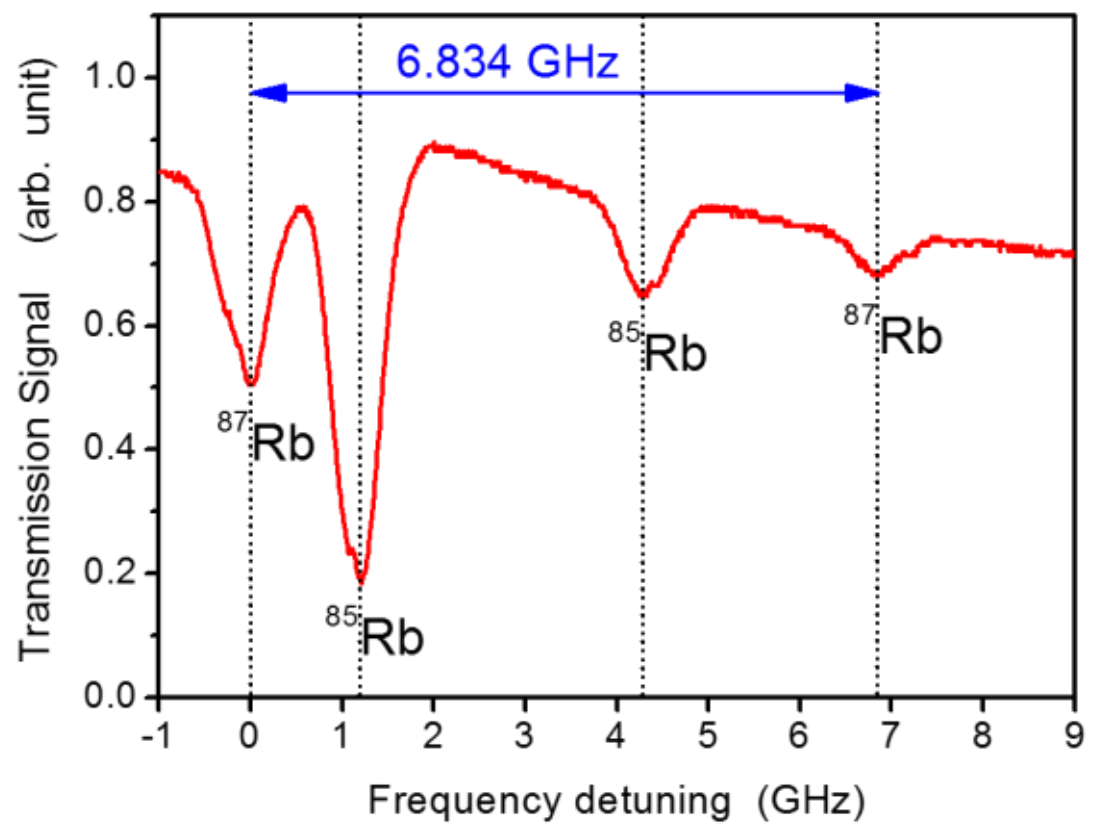

**Figure 4** Doppler-broadened absorption spectra of $5S_{1/2}$ - $5P_{3/2}$ transition (D2 line) of $^{87}$Rb and $^{85}$Rb atoms obtained using 780 nm laser, while the 1560 nm fundamental-wave laser is frequency-scanned linearly. This indicates that the continuous frequency tunable range of the 780 nm laser is at least 10 GHz.

## 4. Single-pass SFG using PPKTP and PPMgO:sLT crystals

Previously, we experimentally presented a three-mirror folded singly-resonant SFG cavity, which was applied to 520 nm single-frequency laser generation via SFG of a 780 nm laser and a 1560 nm laser in a PPKTP crystal [12]. 268 mW of SFG output power was obtained at 520 nm using 6.8 W of 1560 nm laser and 1.5 W of 780 nm laser. However, we want to increase the output power at 520 nm and simplify the entire structure, so the single-pass SFG configuration now is employed.

We use two different crystals for SFG, PPKTP (Raicol Crystals; crystal dimensions of 20 mm×2 mm×1 mm; poling period of 9.1 μm; type 0 matching; both ends of the PPKTP crystal are flat surfaces with anti-reflection coatings for 1560 nm, 780 nm and 520 nm, and the residual reflectivity $R < 0.5\%$) and PPMgO:sLT (HC Photonics; crystal dimensions of 30 mm×1 mm×1 mm; poling period of 7.97 μm; both ends of the PPMgO:sLT crystal are flat surfaces with anti-reflection coatings for 1560 nm, 780 nm and 520 nm, and the residual reflectivity $R < 0.5\%$). First, the PPKTP crystal offers a larger effective nonlinear coefficient, a wider temperature acceptance bandwidth and a lower photo-refractive damage threshold [27], and it is also an attractive quasi-phase-matched (QPM) candidate material for CW green laser generation. Periodically-poled $LiTaO_3$ (PPLT) is the alternative nonlinear conversion material, because of its wide transparency wavelength range (0.28~5.5 μm), high nonlinear coefficients [28], and high resistance to optical damage. Commercially available $LiTaO_3$ is mostly congruent-type material. Recently, stoichiometric $LiTaO_3$ (sLT) crystals have been grown [29]. The coercive field in sLT (1.7~4.5 kV/mm) is lower than that in the congruent-type material. The low coercive field of these stoichiometric crystals is apparently related to their reduced defect density. This low value eases the electrical poling of thicker crystals for use in QPM frequency conversion [30]. Some basic parameters of the PPKTP and PPMgO:sLT bulk crystals that we used in our SFG experiment are listed in Table 1.

**Table 1.** Parameters of PPKTP and PPMgO:sLT crystals used in the experiments

| | Type | Size (mm) | $d_{eff}$ (pm/V) | Poling Period (μm) | Temp. (℃) | ΔT (℃) |
|---|---|---|---|---|---|---|
| PPKTP | 0 | 20x 2 x1 | ~9.5[31] | 9.10 | 57.9 | 1.1 |
| PPMgO:sLT | 0 | 30x 1 x1 | ~6.7[31] | 7.97 | 142.2 | 0.8 |

The PPKTP crystal is placed in a homemade oven, which was made from red copper with a thermal-electric cooler (TEC) and a temperature sensor. The temperature is stabilized using a precision temperature controller. The PPMgO:sLT crystal is placed in a commercial oven from HC Photonics. We can then achieve optimized phase matching by adjusting the temperatures of the crystals. Figure 5 (a) shows the measured temperature tuning data for SFG using the PPKTP crystal. The input powers at 1560.50 nm and 780.25 nm are fixed at ~7 W and ~3 W, respectively. The highest output power at 520.17 nm is obtained when the crystal temperature is tuned to 57.9 ℃. Figure 5 (b) shows the measured temperature tuning data for SFG using the PPMgO:sLT crystal. The measured optimal temperature is 142.2 ℃. The full-width at half-maximum (FWHM) bandwidth values of the phase-matching temperatures of the PPKTP and PPMgO:sLT crystals are ~ 1.1 ℃ and ~ 0.8 ℃, respectively.

However, we cannot simply consider spatial mode matching of the fundamental and doubled beams in SFG crystals; the optimum focusing condition is also important here. Before SFG, the SHG yields ~3 W of 780.25 nm laser output, and the residual fundamental power is ~7 W. Under ideal conditions, according to the consumption of photons, the ratio of power between 1560.50 nm and 780.25 nm is equal to their ratio of frequency (1:2). Here, we must consider the optimum focusing conditions of the doubled beam preferentially, because the available 780 nm laser power is very limited. Figure 6 shows the 520.17 nm green laser output and the optical-optical conversion efficiency versus the input 1560.50 nm laser before SHG, when we use a lens with $f$ = 50 mm as the matching lens. Green dots in Figure 6 (a) and (b) represent the measured data for SFG with the PPKTP bulk crystal. When the 1560.50 nm input power before SHG is 11.64 W, it can yield 545 mW of 520.17 nm green output, corresponding to a maximum optical-optical conversion efficiency of 4.67%, and the normalized nonlinear conversion efficiency is ~ 1.06% $(W{\cdot}cm)^{-1}$. Blue squares in Figure 6 (a) and (b) represent the measured data for SFG using the PPMgO:sLT bulk crystal. When the 1560.50 nm input power before SHG is 11.64 W, it can yield 350 mW of 520.17 nm green laser power, corresponding to a maximum optical-optical conversion efficiency of 3.01%, and the normalized nonlinear conversion efficiency is ~ 0.53% $(W{\cdot}cm)^{-1}$. Other than the optimum focusing condition, the main limiting factor is mismatching of the fundamental and doubled beams in SFG crystals. The focusing lens will lead to color aberration, which means the focusing lengths for the 1560.50 nm and 780.25 nm wavelengths are different. The divergence angles and the waist spot radii in SFG crystals for these two wavelengths are also different. If we focus the 1560.50 nm and 780.25 nm beams separately to match their spatial modes, then the nonlinear conversion efficiency should be improved.

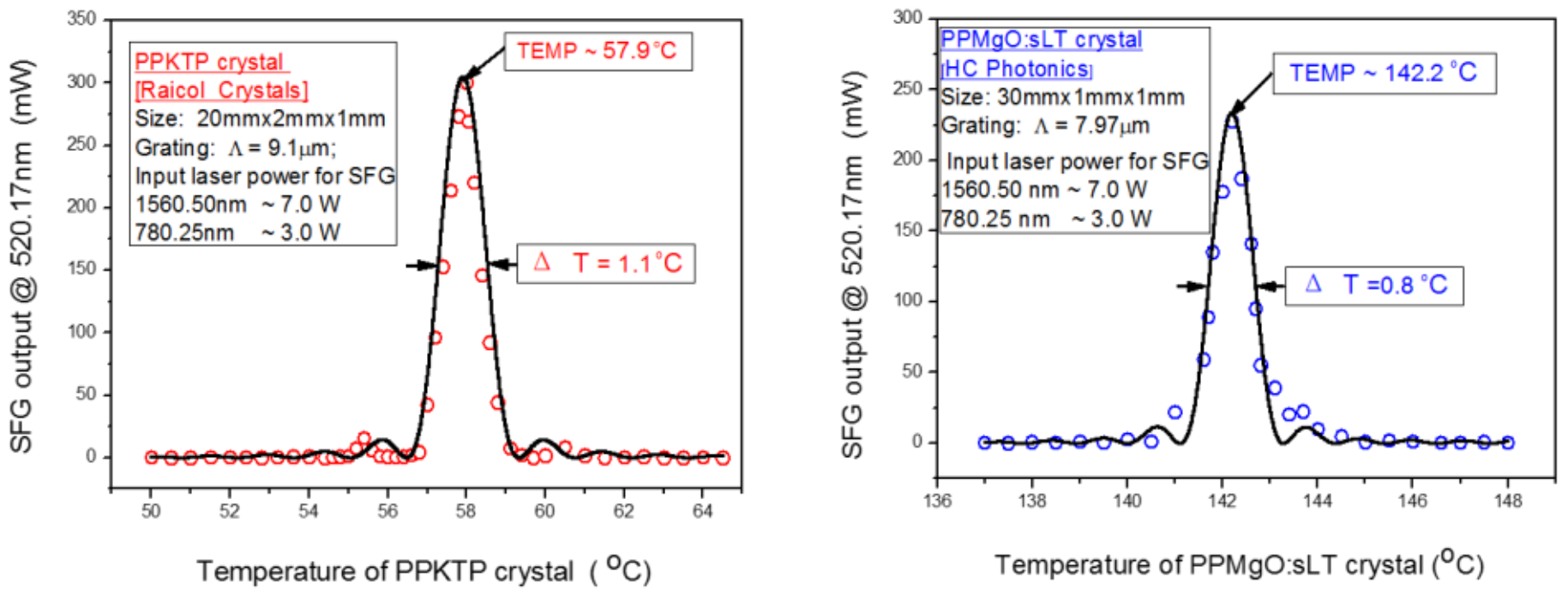

**Figure 5** Temperature tuning curves of PPKTP and PPMgO:sLT crystals for single-pass SFG. Circles represent the experimental data, while the solid lines are theoretically fitted curves using the $sinc^2$ function. (a) PPKTP crystal case, where the quasi-phase-matched temperature is 57.9 ℃ with a FWHM of ~ 1.1 ℃; (b) PPMgO:sLT crystal case, where the quasi-phase-matched temperature is 142.2 ℃ with a FWHM of ~ 0.8 ℃.

We can yield more 520.17 nm laser output using the PPKTP bulk crystal. While the nonlinear coefficient of the PPMgO:sLT bulk crystal is larger than that of the PPKTP crystal, we think that maybe the main factor here is the crystal quality itself, because the beam quality of the 520.17 nm laser for using PPKTP is a bit better than that for using PPMgO:sLT. The transverse beam qualities of the 520.17 nm laser beams obtained using PPKTP and PPMgO:sLT are evaluated using the $M^2$ beam quality factor parameter in the two orthogonal transverse directions $X$ and $Y$. Figure 7 shows the $1/e^2$ beam radius versus the axial position $Z$ after a plano-convex lens with a focal length of 150 mm. For the PPKTP crystal, fitting of the experimental data gives $M_X^2$ = 1.08 and $M_Y^2$ = 1.04. For the PPMgO:sLT crystal, fitting of the experimental data gives $M_X^2$ = 1.16 and $M_Y^2$ = 1.18.

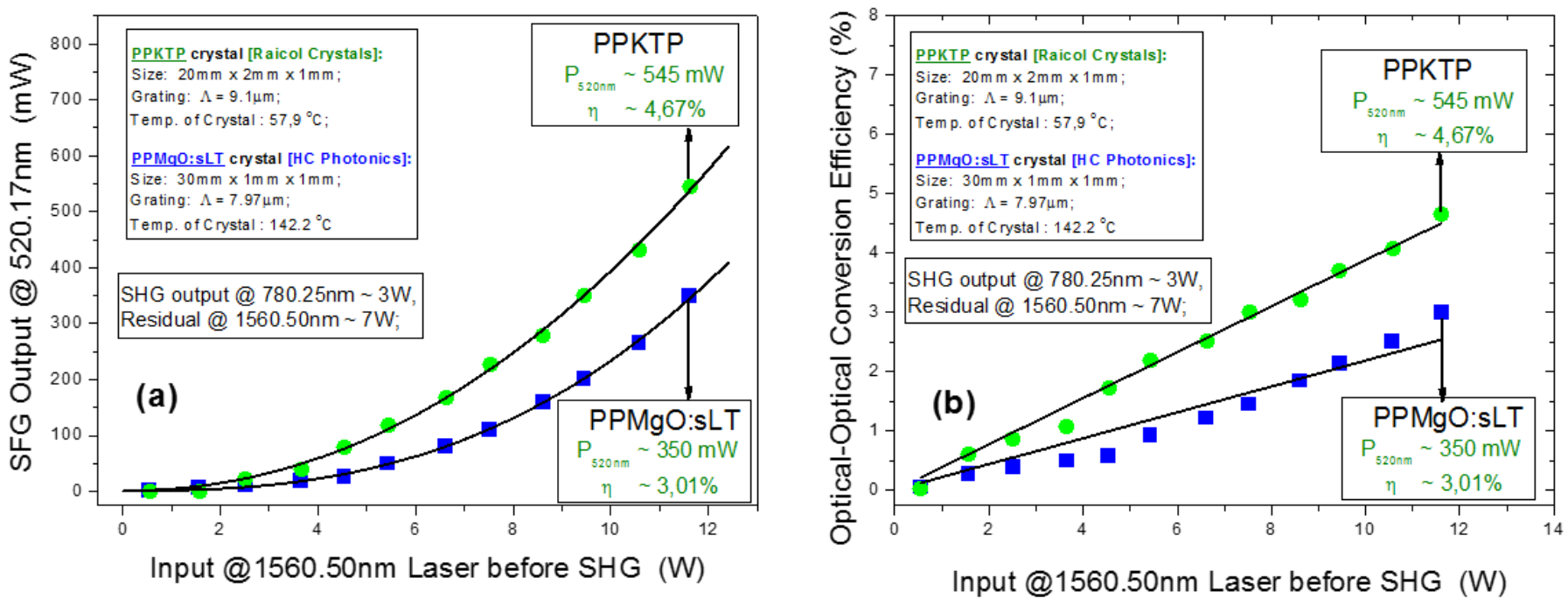

**Figure 6** (a) 520.17 nm green laser output power versus input 1560.50 nm laser power before SHG. (b) Optical-optical conversion efficiency versus input 1560.50 nm laser power before SHG.

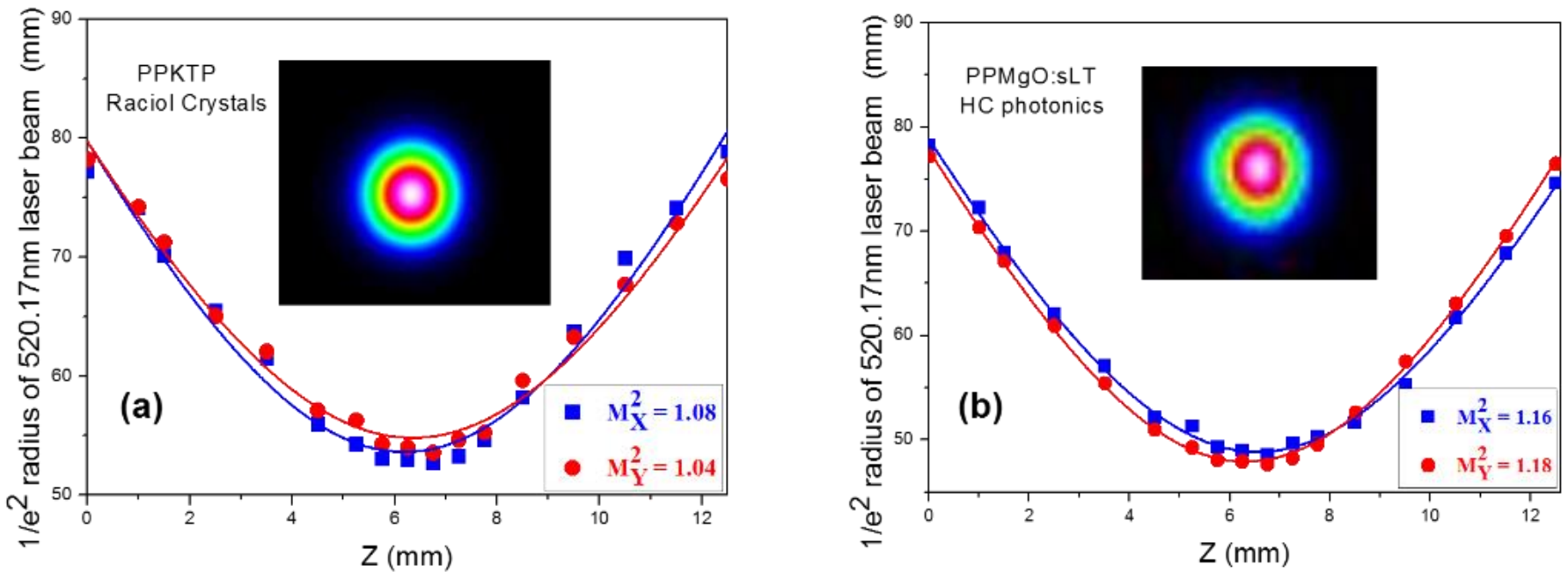

**Figure 7** Beam quality factor ($M^2$) characteristics of the THG laser beam. Insets show the typical intensity profiles of the corresponding THG laser beams. (a) The PPKTP crystal case, where $M_X^2$ = 1.08 and $M_Y^2$ = 1.04. (b) The PPMgO:sLT crystal case, where $M_X^2$ = 1.16 and $M_Y^2$ = 1.18.

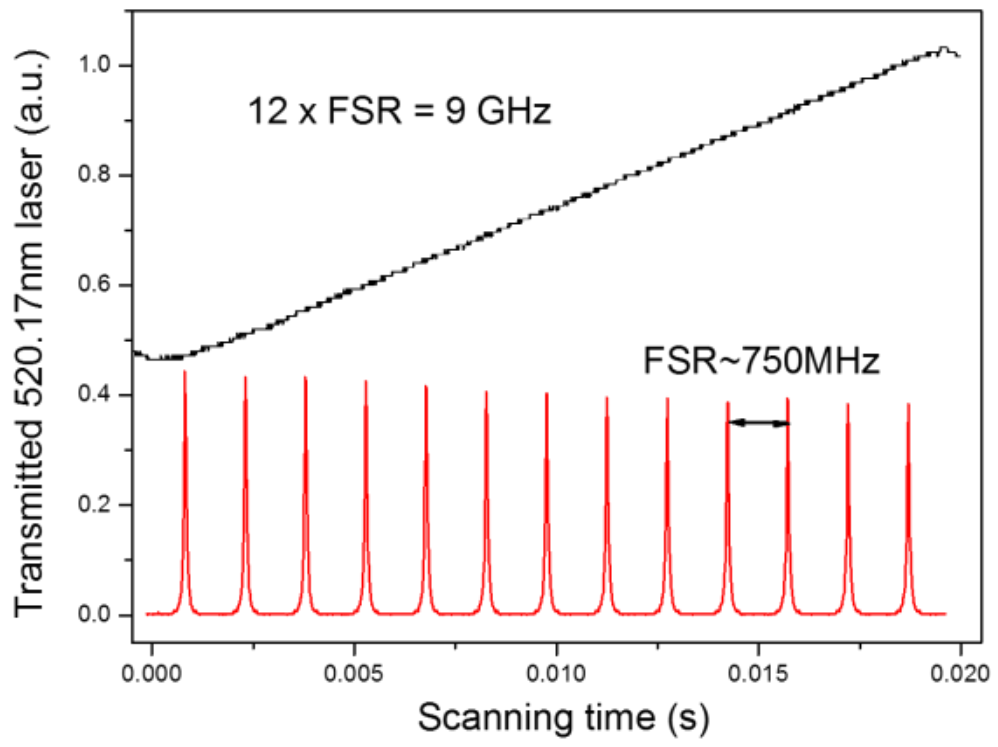

**Figure 8** Scanned 520.17 nm laser beam is monitored using a confocal Fabry-Perot cavity with FSR of 750 MHz. The continuously tunable range of the 520.17 nm laser's frequency is more than 9 GHz.

Finally, we analyze the transmitted signal of a confocal Fabry-Perot cavity with a free spectral range (FSR) of 750 MHz while the 520.17 nm laser frequency is linearly scanned. Figure 8 shows a typical result, where the 520.17 nm laser can be tuned across more than 12 FSRs, which means that the continuously tunable range is more than 9 GHz. Also, we yield the 520.17 nm laser with the single-pass SFG configuration, and thus the tunable range of the 520.17 nm laser is mainly limited by the tunability of the 1560 nm seed laser.

## 5. 520 nm green laser pumped 1560 nm + 780 nm doubly-resonant OPO

In this section, we present a scheme for generation and homodyne detection of two-color entangled continuous-variable optical fields at 1560 and 780 nm from a 520 nm laser-pumped doubly-resonant OPO. Figure 9 shows a schematic diagram of the OPO and the homodyne detection system. The doubly-resonant OPO consists of two plano-concave mirrors with 30 mm radii of curvature and a 20-mm-long temperature-stabilized PPKTP bulk crystal (Raicol Crystals; crystal dimensions of 20 mm×2 mm×1 mm; poling period of 9.1 μm; type 0 matching; both ends of the PPKTP crystal are flat surfaces with anti-reflection coatings for 1560 nm, 780 nm and 520 nm, and the residual reflectivity R < 0.5%). The two cavity mirrors are coated to have high reflectivity at 1560 nm and 780 nm (>99.8%), and high transmission at 520 nm (98%). The OPO cavity is actively locked. Two sets of homodyne detection systems operating at 780 and 1560 nm with corresponding local oscillator beams and radio-frequency spectrum analyzers are also implemented to characterize the quantum noise of the signal and idler beams from the OPO. Here, the 1560 and 780 nm local oscillator beams come from the original 1560 nm output and its frequency doubled output.

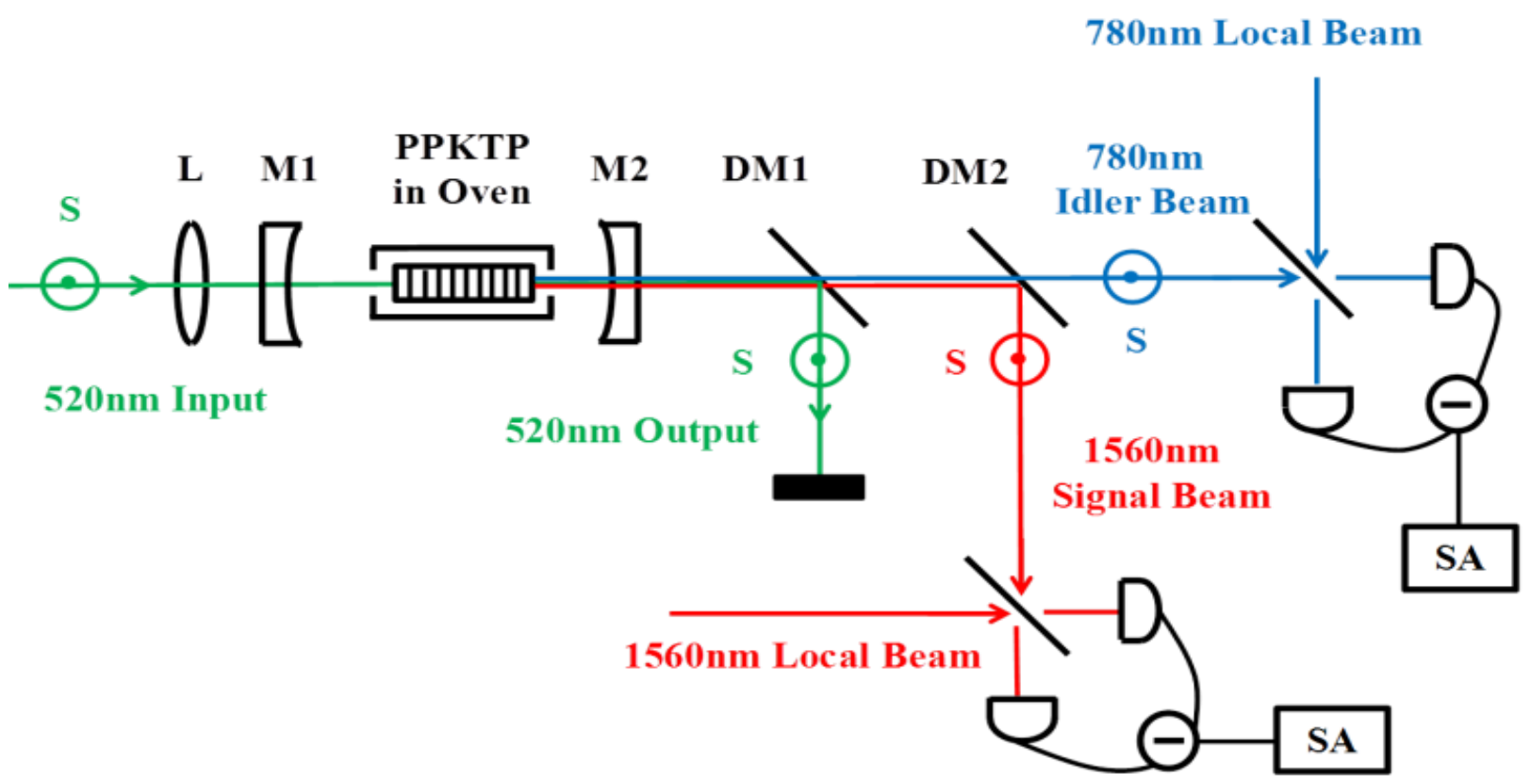

**Figure 9** Schematic diagram of OPO and homodyne detection system. Ms: plano-concave OPO cavity mirrors; DMs: dichroic mirrors; SA: radio-frequency spectrum analyzer. The 520 nm, 780 nm, and 1560 nm beams are all s-polarized.

We achieved coarse tuning of both the signal and idler beams by varying the temperature of the PPKTP crystal. Figure 10 shows the relationships between the signal and idler wavelengths and the crystal temperature. The solid dots and solid squares indicate the data for the signal and idler beams, respectively. When the crystal temperature is adjusted from 26 to 80 ℃, the signal wavelength can be coarsely tuned from 1529.81 nm to 1573.83 nm; correspondingly, the idler wavelength can be coarsely tuned from 788.26 nm to 777.20 nm. Specifically, when the crystal temperature is precisely stabilized at 65.0 ℃, we can achieve both 1560.50 nm and 780.25 nm outputs. Additionally, the 780.25 nm idler beam can be continuously scanned across the Rb D2 line by scanning the 520 nm pump laser’s frequency. Further experiments are now being undertaken to verify and characterize the two-color quantum entanglement between the signal and idler beams.

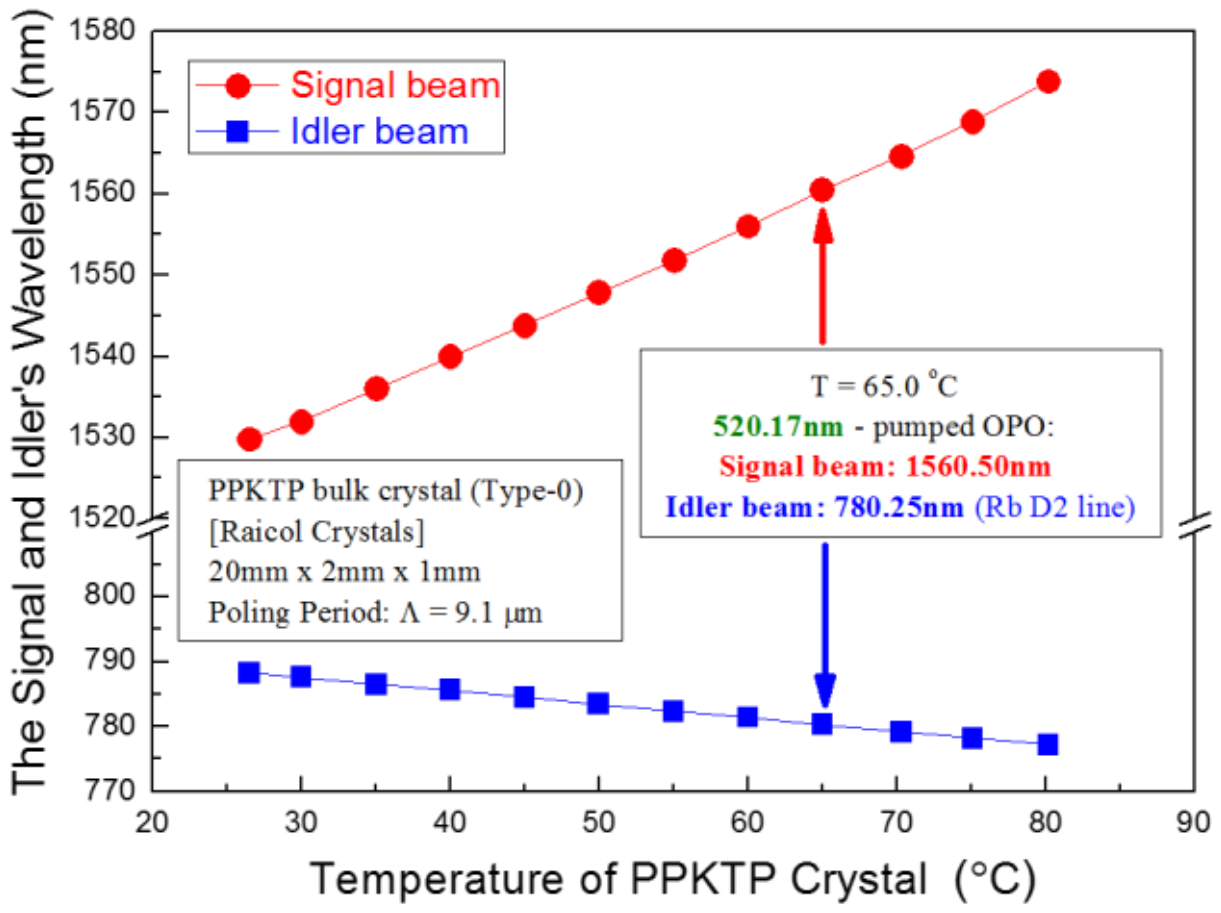

**Figure 10** Coarse tunability of the signal and idler beams from the OPO indicated by the signal and idler wavelengths versus the PPKTP crystal's temperature. As the crystal temperature changes from 26 to 80 ℃, the signal wavelength can be coarsely tuned from 1529.81 nm to 1573.83 nm (~44 nm), while the corresponding idler wavelength can be coarsely tuned from 788.26 nm to 777.20 nm (~11 nm).

## 6. Conclusion and perspectives

We have demonstrated a simple, compact and efficient laser frequency generation scheme that combines single-pass SHG and single-pass SFG. To increase the SHG output power, we used two cascaded PPMgO:LN crystals, and demonstrated 3.53 W of 780.25 nm single-frequency continuous-wave laser output from a 14 W EDFA-boosted 1560 nm ECDL, where the measured maximum doubling efficiency is 26.8%. We can attempt to use more crystals or increase the length of the single crystal to yield more 780 nm output.

We then compared single-pass SFG using PPMgO:sLT and PPKTP crystals. When the 1560.50 nm laser power before SHG is 11.64 W, it can yield 545 mW of 520.17 nm green light power when using the PPKTP crystal; the same conditions can also yield 350 mW of 520.17 nm green laser power when using the PPMgO:sLT crystal. If we use 50-mm-long crystals, more green output should be expected reasonably. The single-frequency moderate-output 520 nm green laser has a high beam quality, and can be continuously tuned over a very wide frequency range. Specifically, because the 520.17 nm and 780.25 nm laser beams are generated using the 1560.50 nm source, the system is thus coherent, and can then provide 780.25 nm and 1560.50 nm local oscillators for implementation of homodyne detection to verify the two-color quantum entanglement. This task cannot be addressed using a direct 520.17 nm green laser.

## Acknowledgments

This project is supported by the National Natural Science Foundation of China (Grant Nos. 11274213, 61475091, and 61227902), and the National Major Scientific Research Program of China (Grant No. 2012CB921601). Thanks Shanlong GUO and Yulong Ge's contributions for some basic works related to this subject. Also thanks Prof. Yongmin LI in our institute for loan of some specific mirrors.

## References

1. F. Lienhart, S. Boussen, O. Carraz, N. Zahzam, Y. Bidel, and A. Bresson, "Compact and robust laser system for rubidium laser cooling based on the frequency doubling of a fiber bench at 1560 nm," Appl. Phys. B **89** 177-180 (2007).
2. J. Dingjan, B. Darquie, J. Beugnon, M.P.A. Jones, S. Bergamini, G. Messin, A. Browaeys, and P. Grangier, "A frequency-doubled laser system producing ns pulses for rubidium manipulation," Appl. Phys. B **82** 47-51 (2006).
3. Y. Sortais, S. Bize, C. Nicolas, C. Salomon, and C. Willams, "Cold Collision Frequency Shifts in a $^{87}$Rb Atomic Fountain," Phys. Rev. Lett. **85** 3117-3120 (2000).
4. M. D. Eisaman, A. Andre, F. Massou, M. Fleischhauer, A. S. Zibrov, and M. D. Lukin, "Electromagnetically induced transparency with tunable single-photon pulses," Nature **438** 837-841 (2005).
5. O. Carraz, F. Lienhart, R. Charriere, M. Cadoret, N. Zahzam, and A. Bresson, "Compact and robust laser system for onboard atom interferometry," Appl. Phys. B **97** 405-411 (2009).
6. Chunchun Liu, Xiaomin Guo, Zengliang Bai, Xuyang Wang, and Yongmin Li, "High-efficiency continuously tunable single-frequency doubly resonant optical parametric oscillator," Appl. Opt. **50** 1477-1481 (2011).
7. Shanlong Guo, Yulong Ge, Kong Zhang, Jun He, and Junmin Wang, "Doubly-resonant 780 nm + 1560 nm Optical Parametric Oscillator Pumped by 520 nm Laser," Acta Sinica Quantum Optica **21** 93-98 (2015) (in Chinese).
8. Shuqin Zhai, Rongguo, Yang, Kui Liu, Hailong, Zhang, Junxiang, Zhang, and Jiangrui Gao, "Bright two-color tripartite entanglement with second harmonic generation," Opt. Express **17** 9851-9857 (2009).
9. A. S. Villar, L. S. Cruz, K. N. Cassemiro, M. Martinelli, and P. Nussenzveig, "Generation of bright two-color continuous variable entanglement." Phys. Rev. Lett. **95** 243603 (2005).

10. Xiaomin Guo, Changde Xie, and Yongmin Li, "Generation and homodyne detection of continuous-variable entangled optical beams with a large wavelength difference," Phys. Rev. A. **84** 020301 (2011).
11. Xiaowei Deng, Shuhong Hao, Hong Guo, Changde Xie, and Xiaolong Su, "Continuous variable quantum optical simulation for time evolution of quantum harmonic oscillators," Sci. Report **6** 22914 (2016).
12. Shanlong Guo, Yulong Ge, Jun He, and Junmin Wang, "Singly resonant sum-frequency generation of 520-nm laser via a variable input-coupling transmission cavity," J. Mod. Opt. **62** 1583-1590 (2015).
13. S. Vasilyev, A. Nevsky, I. Ernsting, M. Hansen, J. Shen, and S. Schiller, "Compact all-solid-state continuous-wave single-frequency UV source with frequency stabilization for laser cooling of Be+ ions," Appl. Phys. B **103** 27-33 (2011).
14. C. Philippe, E. Chea, Y. Nishida, F. du Burck, and O. Acef, "Efficient third harmonic generation of a CW-fibered 1.5 μm laser diode," Appl. Phys. B **122** 265 (2016).
15. Qiang Hao, Qingshan Zhang, Tingting Sun, Jie Chen, Zhanhua Guo, Yuqing Wang, Zhengru Guo, Kangwen Yang, and Heping Zeng, "Divided-pulse nonlinear amplification and simultaneous compression," Applied Physics Letters. **106** (10) 101103 (2016).
16. Qiang Hao, Yunfeng Wang, Tingting Liu, Hong Hu, and Heping Zeng "Divided-Pulse Nonlinear Amplification at 1.5 μm," IEEE Photonics Journal. **8** (5) 1-8 (2016).
17. R. J. Thompson, M. Tu, D. C. Aveline, N. Lundbland, and L. Maleki, "High power single frequency 780nm laser source generated from frequency doubling of a seeded fiber amplifier in a cascade of PPLN crystals," Opt. Express. 111709-1713 (2003).
18. R. Geiger, V. Ménoret, G. Stern, N. Zahzam, P. Cheinet, B. Battelier, A. Villing, F. Moron, M. Lours, Y. Bidel, A. Bresson, A. Landragin, and P. Bouyer, "Detecting inertial effects with airborne matter-wave interferometry," Nature Commun. **2** 474 (2011).
19. M. I. Hussain, M. J. Petrasiunas, C. D. B. Bentley, R. L. Taylor, A. R. R. Carvalho, J. J. Hope, E. W. Streed, M. Lobino, and D. Kielpinski, "Ultrafast, high repetition rate, ultraviolet, fiber laser source: application towards $Yb^+$ fast quantum-logic," Opt. Express **24** 16638-16648 (2016).
20. Bo Xu, Zhanyu Hou, Mingjun Xia, Lijuan Liu, Xiaoyang Wang, Rukang Li, and Chuangtian Chen, "High average power third harmonic generation at 355 nm with $K_3B_6O_{10}Br$ crystal," Opt. Express

**24** 10345-10351 (2016).
21. S. Chiow, T. Kovachy, J. M. Hogan, and M. A. Kasevich, "Generation of 43 W of quasi-continuous 780 nm laser light via high-efficiency, single-pass frequency doubling in periodically poled lithium niobate crystals," Opt. Lett. **37** 3861-3863 (2012).
22. G. D. Boyd and D. A. Kleinman, "Parametric interaction of focused Gaussian light beams," J. Appl. Phys. **39** 3597-3639 (1968).
23. S. C. Kumar, G. K. Samanta, K. Devi, and M. Ebrahim-Zadeh, "High-efficiency, multicrystal, single-pass, continuous-wave second harmonic generation," Opt. Express **19** 11152-11169 (2011).
24. K. Devi, S. Parsa, and M. Ebrahim-Zadeh, "High-efficiency, multicrystal, single-pass, continuous-wave second harmonic generation," Opt. Express **24** 8763-8769 (2016).
25. A. K. Hansen, O. B. Jensen, B. Sumpf, G. Erbert, A. Unterhuber, W. Drexler, P. E. Andersen, and P. M. Petersen, "Generation of 3.5 W of diffraction-limited green light from SHG of a single tapered diode laser in a cascade of nonlinear crystals," Proc. SPIE **8964** 896406 (2014).
26. Yulong Ge, Shanlong Guo, Yashuai Han, and Junmin Wang, "Realization of 1.5 W 780nm single-frequency laser by using cavity-enhanced frequency doubling of an EDFA boosted 1560nm diode laser," Opt. Commun. **334** 74-78 (2015).
27. Mei Sang, Quasi-phase matched PPKTP fabrication and relative application research (Tianjin University, doctoral dissertation, 2003).
28. Shining Zhu, Yongyuan Zhu, Haifeng Wang, Zhiyong Zhang, Naiben Ming, Wenzong Shen, Yong Chang and Xuechu Shen, "Second-order quasi-phase-matched blue light generation in a bulk periodically poled $LiTaO_3$," J. Phys. D: Appl. Phys. **28** 2389-2391 (1995).
29. T. Hatanaka, K. Nakamura, T. Taniuchi, H. Ito, Y. Furukawa, and K. Kitamura, "Quasi-phase-matched optical parametric oscillation with periodically poled stoichiometric $LiTaO_3$," Opt. Lett. **25** 651-653 (2000).
30. Yanhua Lu, Lei Zhang, Yi Ma, Dong Liu, Chun Tang, Weimin Wang, Songxin Gao, and Bin Wei, "Sodium Guidestar Laser Based on High-Efficiency PPSLT Quasi-Phase-Matched Sum Frequency Generation," Acta Optica Sinica **30** 2306-2310 (2010) (in Chinese).
31. S. Manjoorana, H. Zhaoa, I. T. Lima Jr. b, and A. Majora, "Phase-matching properties of PPKTP, MgO: PPSLT and MgO: PPcLN for ultrafast optical parametric oscillation in the visible and near-infrared ranges with green pump," Laser Physics **22**(8) 1325-1330 (2012).